\documentclass[aps,prb,twocolumn,showpacs,superscriptaddress]{revtex4}

\usepackage{graphicx} 
\usepackage{soul}
\begin{document}

\title{Two and One-dimensional Honeycomb Structure of Boron Nitride}

\author{M. Topsakal}
\affiliation{UNAM-Institute of Materials Science and
Nanotechnology, Bilkent University, Ankara 06800, Turkey}
\author{E. Akt\"{u}rk}
\affiliation{UNAM-Institute of Materials Science and
Nanotechnology, Bilkent University, Ankara 06800, Turkey}
\author{S. Ciraci}\email{ciraci@fen.bilkent.edu.tr}
\affiliation{UNAM-Institute of Materials Science and
Nanotechnology, Bilkent University, Ankara 06800, Turkey}
\affiliation{Department of Physics, Bilkent University, Ankara
06800, Turkey}

\date{\today}

\begin{abstract}
This paper presents a systematic study of two and one dimensional
honeycomb structure of boron nitride (BN) using first-principles
plane wave method.  Two-dimensional (2D) graphene like BN is a wide band gap
semiconductor with ionic bonding. Phonon dispersion curves demonstrate the
stability of 2D BN flakes. Quasi 1D armchair BN nanoribbon are
nonmagnetic semiconductors with edge states. Upon passivation of B
and N  with hydrogen atoms these edge states disappear and band gap increases. Bare
zigzag BN nanoribbons are metallic, but become a ferromagnetic
semiconductor when their both edges are passivated with hydrogen.
However, their magnetic ground state, electronic band structure
and band gap are found to be strongly dependent on whether B- or
N-edge of the ribbon is saturated with hydrogen. Vacancy defects
in armchair and zigzag nanoribbons affects also magnetic state and
electronic structure. In order to reveal dimensionality
effects these properties are contrasted with those of various 3D BN
crystals and 1D BN atomic chain.
\end{abstract}


\maketitle

\section{introduction}
Synthesis of a single atomic plane of graphite, i.e.
\textit{Graphene} with covalently bonded honeycomb lattice has
been a breakthrough for several reasons \cite{novo,zhang,berger}.
Firstly, electrons behaving as if massless Dirac Fermions have
made the observation of several relativistic effects possible.
Secondly, stable graphene has disproved previous theories, which
were concluded that two-dimensional structures cannot be stable.
Graphene displaying exceptional properties, such as high mobility
even at room temperature, ambipolar effect, Klein tunelling, anomalous  
quantum hall effect  etc. seems to offer
novel applications in various fields \cite{graphene_applications1}.
Not only 2D graphene, but also its quasi 1D forms, such as
armchair and zigzag nanoribbons have shown novel electronic and
magnetic properties \cite{graphene_applications2,graphene_applications3,graphene_applications4}, 
which can lead to important applications in
nanotechnology. As a result, 2D honeycomb structures derived from
Group IV elements and Group III-V and II-VI compounds are
currently generating significant interest owing to their unique
properties.

Boron-Nitride (BN) in ionic honeycomb lattice which is the Group
III-V analogue of graphene have also been produced having desired
insulator characteristics \cite{bn-synthesis}.
Nanosheets \cite{bn-nanosheets1,bn-nanosheets2}, nanocones \cite{bn-nanocones}, 
nanotubes \cite{bn-nanotubes}, nanohorns \cite{bn-nanohorns}, 
nanorods \cite{bn-nanorods} and nanowires \cite{bn-nanowires} of
BN have already been synthesized and these systems might hold
promise for novel technological
applications. Among all these different
structures, BN nanoribbons, where the charge carriers are confined
in two dimension and free to move in third direction, are
particularly important due to their well defined geometry and
possible ease of manipulation.

BN nanoribbons  posses different electronic and magnetic 
properties depending on their size and edge termination. Recently,
the variation of band gaps of BN nanoribbons with their widths and
Stark effect due to applied electric field have been
studied \cite{Guo,louie-nanoletter}. Recently the magnetic properties of zigzag
BN nanoribbons have been investigated \cite{barone}. Half-metallic properties 
have been revealed from these studies which
might be important for spintronic applications. Production of
graphene nanoribbons as small as 10 nm in width has been achieved
\cite{dai1,dai2} and similar techniques are expected to be
developed for BN nanoribbons.

A thorough understanding of 2D BN honeycomb structure and their
various nanoribbons is important for further study of this graphene
like compounds. BN by itself provides with very interesting
chemical and physical properties, which may lead to important
applications. In this paper, we present a
detailed ab-initio study of electronic, magnetic and elastic
properties of 2D (graphene like) BN and bare and hydrogen
passivated, quasi 1D BN nanoribbons (BNNRs). We also investigated
the effect of the vacancy defects on these properties. To reveal
the dimensionality effects we include also a short discussion
regarding 3D BN bulk crystals and 1D BN atomic chains. We found
that 2D BN is a nonmagnetic, wide band gap semiconductor. The
ionic bonding due to significant amount of charge from B to N atom
opens a gap and hence dominates electronic structure. Calculated
phonon dispersion curves provide a clear evidence that 2D BN flakes
is stable. The armchair and zigzag nanoribbons of BN display even
more interesting electronic and magnetic properties. Bare and hydrogen passivated
armchair BN nanoribbons (A-BNNR) are nonmagnetic wide band gap
semiconductor. The value of band gap of A-BNNR having width $w >
10$ \AA~ is practically independent from the width of nanoribbons.
While the bare zigzag BN nanoribbons (Z-BNNR) are ferromagnetic
metal, they become nonmagnetic semiconductor upon the passivation
of both edges. We found that 2D BN and its nanoribbons have
properties, which are complementary to graphene.

\section{Model and Methodology}

We have performed first-principles plane wave calculations within
density functional theory (DFT) using PAW potentials \cite{paw}.
The exchange correlation potential has been approximated by
generalized gradient approximation (GGA) using PW91 \cite{pw91}
functional both for spin-polarized and spin-unpolarized cases. All
structures have been treated within supercell geometry using the
periodic boundary conditions. A plane-wave basis set with kinetic
energy cutoff of 500 eV has been used. In the self-consistent
potential and total energy calculations the Brillouin zone (BZ) is
sampled by special \textbf{k}-points. The numbers of these
\textbf{k}-points are (15x15x15) for bulk BN, (25x25x1) for 2D BN
and (25x1x1) for nanoribbons, respectively, and are scaled
according to the size of superlattices. All atomic positions and
lattice constants are optimized by using the conjugate gradient
method, where the total energy and atomic forces are minimized.
The convergence for energy is chosen as 10$^{-5}$ eV between two
steps, and the maximum Hellmann-Feynman forces acting on each atom
is less than 0.02 eV /\AA~upon ionic relaxation. A large spacing
($\sim$ 10 \AA) between monolayers has been taken to prevent
interactions between them. The pseudopotentials having 3 and 5
valence electrons for the B (B: $2s^{2}$ $2p^{1}$) and N ions (N:
$2s^{2}$ $2p^{3}$) were used. Numerical calculations have been
performed by using VASP package \cite{vasp1,vasp2}. The phonon
dispersion curves are calculated within density
functional perturbation theory (DFPT) using plane wave methods as
implemented in PWSCF software \cite{pwscf}.

\section{3D  BN Crystals and 1D atomic chain}

In this section, we present our theoretical calculations on 3D
bulk BN crystals and truly 1D BN atomic chain. Earlier these 3D
bulk crystals \cite{BNcrystal1,BNcrystal2,BNcrystal3,BNcrystal4,BNcrystal5} 
and 1D atomic chains \cite{BNchain} have been studied theoretically
by using different methods. Our purpose
in including these crystals of BN in different dimensionalities is
to contrast their properties with those of 2D and quasi 1D
honeycomb structures of BN and also reveal dimensionality effects.

\subsection{3D Bulk BN crystals}
\begin{widetext}
\begin{figure*}
\begin{center}
\includegraphics[width=17cm]{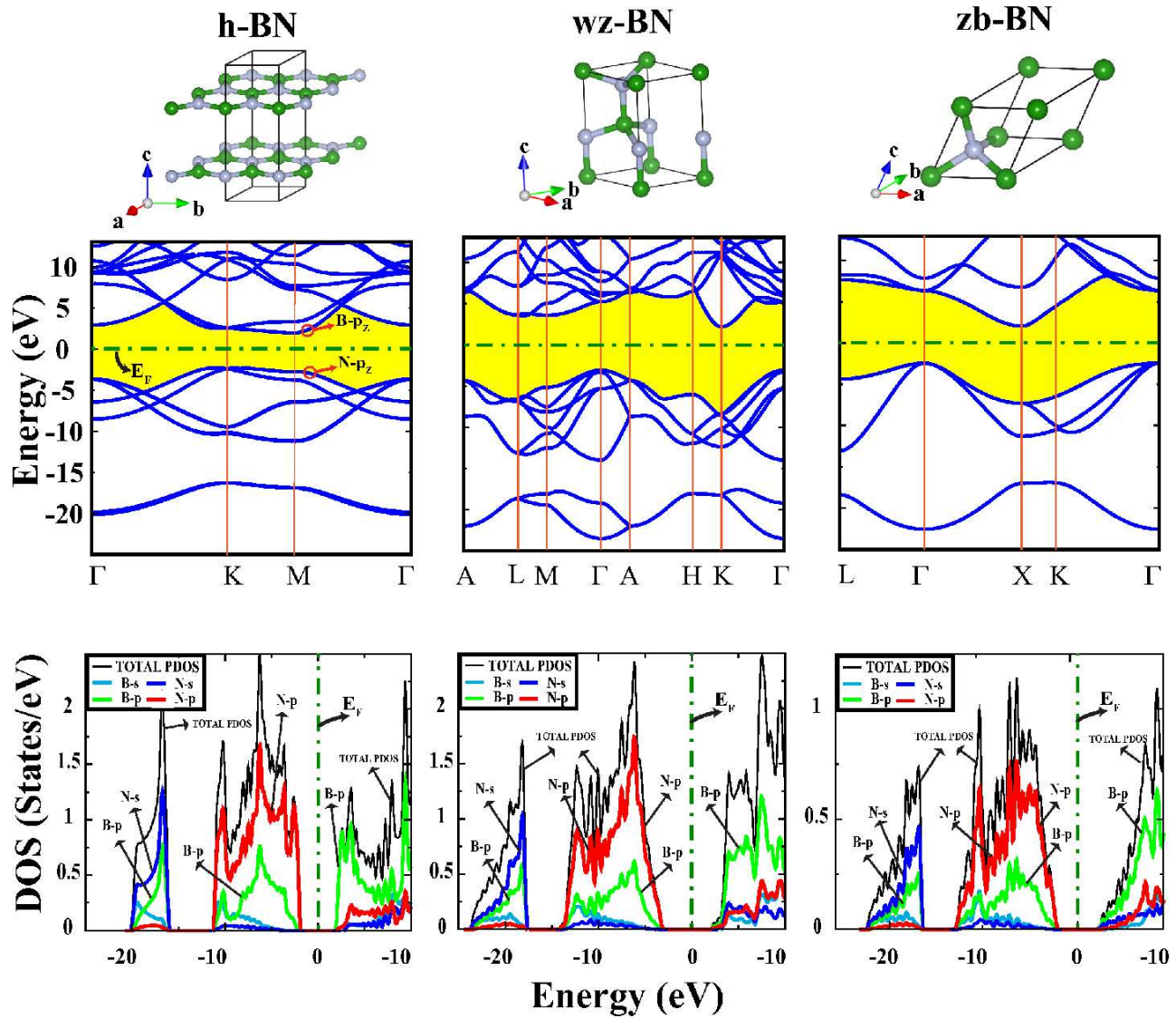}
\caption{(Color online) Optimized atomic structure, energy bands,
total (TDOS) and orbital projected density of states (PDOS) of
various 3D crystals of BN. (a) Hexagonal (h-BN) whose B(N) atoms are 
on top of the N(B) atoms in the consecutive layer; (b) wurtzite (wz-BN);
and (c) zincblende (zb-BN) crystals. Dark-green and light-gray balls
represent B and N atoms, respectively. The band gaps between
conduction and valence bands are highlighted. The orbital
character of states are indicated for the conduction and valence
band edges. The zero of energy is set to the Fermi energy E$_{F}$.
All structures are fully optimized.} \label{fig:Figure-Bulks}
\end{center}
\end{figure*}
\end{widetext}

Three dimensional bulk crystals include hexagonal-layered BN
(h-BN), wurtzite BN (wz-BN) and zincblende BN (zb-BN) structures.
Their atomic configurations and primitive unit cells are described
in Fig.~\ref{fig:Figure-Bulks}. By using the expression,

\begin{equation}\label{equ:binding}
E_C={E[BN]} - E[B] - E[N]
\end{equation}
where E[BN] is the total energy per B-N pair of the optimized
structure of BN crystal; E[B] and E[N] are the total energies of
free B and N atoms; we calculated the equilibrium cohesive
energies of h-BN, wz-BN and zb-BN crystals as -17.65, -17.45 and
-17.49 eV per B-N pair, respectively. Accordingly, h-BN, which is
the analogue of graphite, is the most energetic bulk structure. On the other hand,
the cubic BN structure is known to be the second hardest material of all.

The lattice constants of the optimized structures in equilibrium
are $a = 2.511$ \AA, $c/a = 2.66$ and the distance between the
nearest B and N atoms is $d=1.450$ \AA for h-BN layered crystal.
For wz-BN, optimized values of $a$, $c/a$ and $d$ are calculated
to be 2.542 \AA, 1.64, and 1.561 \AA, respectively. The zincblende structure has lattice constant $a = 2.561$ \AA~ and
$d=1.568$ \AA. All our results related with the structural
parameters are in good agreement with the experimental and theoretical values
\cite{BNcrystal1,BNcrystal2,BNcrystal3,BNcrystal4,bulk-references} 
within the average error of  $\sim 1\%$.

The calculated electronic band structure, total and partial (or
orbital projected) density of states (DOS) of 3D crystals are
presented in Fig.~\ref{fig:Figure-Bulks}. These h-BN, wz-BN and
zb-BN crystals are indirect band gap semiconductors with
calculated band gaps being $E_{G}$=4.47, 5.72, and 4.50 eV,
respectively. The calculated values of $E_{G}$ differ from the 
earlier ones depending on the method used \cite{DFTgap}. For h-BN having 2D BN atomic layers in the (x,y)-plane. The band structure is composed from the band structures of these individual atomic layers with hexagonal symmetry, which are slightly split due to weak coupling between them. Highest 
valence band has N-$p_{z}$ character; the states of the lowest 
conduction band is formed from B-$p_{z}$ orbitals (the z direction corresponds to ``c'' in Fig.~\ref{fig:Figure-Bulks} ). Overall features of the total density of
states (TDOS) are similar for three 3D crystal structures. Valence
band consists of two parts separated by a wide intra band gap. The
lower part at $\sim$ -20 eV is projected mainly to N-$s$ and
partly to N-$p$ and B-$s$ orbitals. The upper part is due to
mainly N-$p$ and partly B-$p$ orbitals and has similarities in
both zb-BN and wz-BN crystals. As for the lower part of the
conduction band it is derived mainly from B-$p$ orbitals. The
differences of three 3D crystals are pronounced in the lower part
of the conduction band.

We calculate the amount of charge on constituent B and N atoms in
3D crystals by performing the L\"{o}wdin \cite{lowdin} analysis in terms of the
projection of plane-waves into atomic orbitals. By subtracting the
valencies of free B and N atoms from the calculated charge values
on the same atoms in 3D crystals we obtain the charge transfer,
$\Delta Q$. The calculated values of $\Delta Q$ for h-BN, wz-BN
and zb-BN are 0.416, 0.342, 0.334 electrons, respectively. The fact
that $\Delta Q$ of zb-BN and wz-BN have almost equal values, but $\Delta
Q$ of h-BN crystal is significantly larger related to the shorter
B-N bond length in h-BN crystal.

\subsection{1D BN Atomic chain}

BN forms stable segments of linear atomic chain \cite{BNchain} like
carbon \cite{tongay}. This situation is in contrast to second and
third row elements (such as Si and Ge) and III-V compounds and metals
(such as Al, Au etc) which can form stable zigzag chain structures
instead of linear chain structures. Our results on 
optimized chain structure yield the cohesive energy $E_{C}$=16.04
eV per B-N pair, the B-N distance $d=$1.307 \AA, the indirect band
gap E$_{G}=$3.99 eV and charge transfer from B to N, $\Delta
Q=$0.511 electrons. Hence the double bond between B and N is
ionic.

\section{2D Honeycomb Structure of BN}
Having discussed the overall structural and elastic properties of 3D
and 1D BN, we now consider 2D BN with hexagonal symmetry.
The atomic structure of 2D BN is similar to the honeycomb
structure of graphene, except that the constituent atoms of the
former are from III and V columns of the Periodic Table. Normally,
the bond between nearest B and N atoms is formed from the bonding
combination of B-$sp^{2}$ and N-$sp^{2}$ orbitals. However, owing
to the electronegativity difference between B and N atoms
electrons are transferred from B to N. As a result, in contrast to
purely covalent bond in graphene the bonding between B and N gains
an ionic character. The charge transfer from B to N dominates
several properties of 2D BN including the opening of the band gap.
In this respect the BN honeycomb structure is complementary to
graphene.

\subsection{Charge density analysis and electronic structure}

\begin{figure}
\begin{center}
\includegraphics[width=7.5cm]{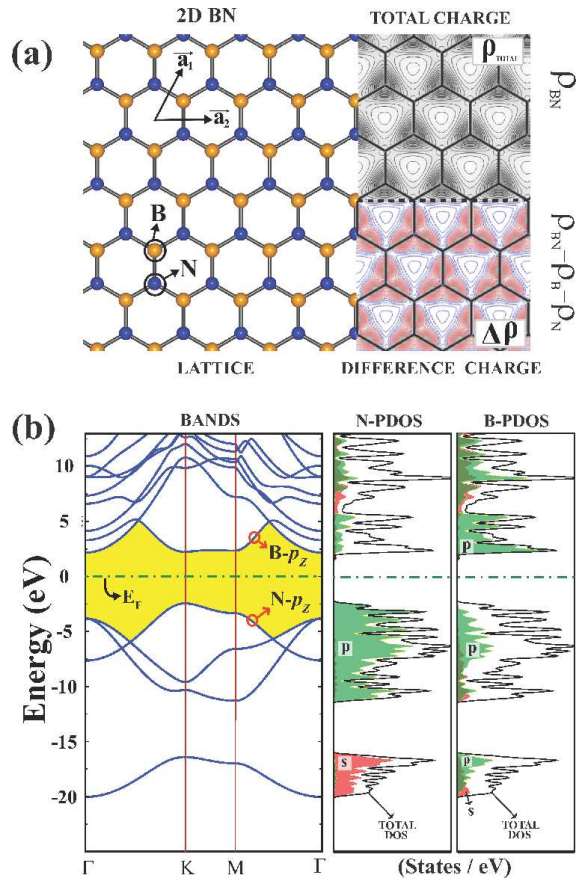}
\caption{(Color online) (a) Primitive unit cell of the honeycomb
structure of 2D BN together with Bravais lattice vectors.
Calculated total charge density $\rho_{BN}$ and difference charge
density $\Delta \rho$, are also shown in the same panel. (b)
Calculated electronic structure of 2D BN honeycomb crystal
together with total, TDOS and partial density of states, PDOS on B
and N atoms. The orbital character of the states are also
indicated.} \label{fig:Figure-ChargeAnaliz}
\end{center}
\end{figure}

The atomic structure, atomic charge, charge transfer from B to N
and the electronic structure of 2D BN are presented in
Fig.~\ref{fig:Figure-ChargeAnaliz}. Contour plots of total charge
indicates high density around N atoms. The difference charge
density is calculated by subtracting charge densities of free B
and N atoms from the charge density of 2D BN, i.e. $\Delta
\rho=\rho_{BN}-\rho_{B}-\rho_{N}$. High density contour plots
around N atoms protruding towards the B-N bonds indicate charge
transfer from B to N atoms. This way the B-N bonds achieve an
ionic character. The amount of transfer of charge is calculated by
L\"{o}wdin analysis to be $\Delta Q$=0.429 electrons.
Interestingly, $\Delta Q$ is slightly larger than that calculated
for  h-BN, but significant larger than those calculated for 
wz-BN and zb-BN crystals.

2D BN is a semiconductor. Calculated electronic energy bands 
are similar to those calculated
for  h-BN crystal. 
The $\pi$- and $\pi^{*}$- bands of graphene
which cross at the K- and K$^{*}$-points of the BZ open a gap in
2D BN as a bonding and antibonding combination of N-$p_{z}$ and
B-$p_{z}$ orbitals. The contribution of N-$p_{z}$ is pronounced
for the filled band at the edge of valence band. The calculated
band gap is indirect and $E_{G}=$4.64 eV. TDOS and partial
density of states show also similarity to those of h-BN layered
crystal presented in Fig.~\ref{fig:Figure-Bulks}.

\subsection{Phonon spectrum}

\begin{figure}
\begin{center}
\includegraphics[width=8cm]{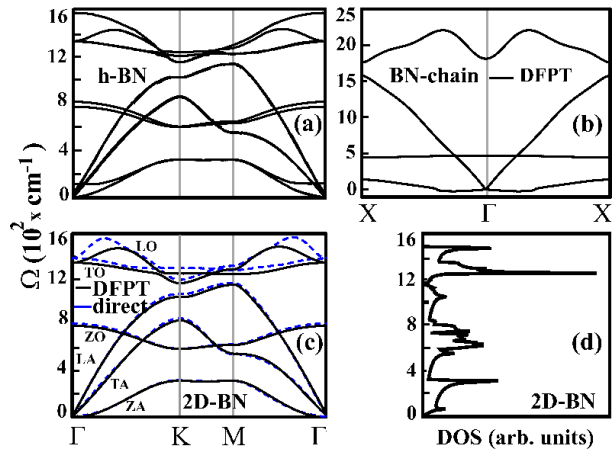}
\caption{(Color online) Calculated phonon frequencies versus
\textbf{k}-vectors. (a) h-BN crystal. (b) 1D BN atomic chain. (c)
2D BN honeycomb structure. Phonon modes calculated by force
constant direct method are shown by the blue-dashed curve. (d) 
Density of phonon frequencies (DOS) for the 2D BN honeycomb
structure.} \label{fig:Figure-Phonon}
\end{center}
\end{figure}

Even if the structure optimization resulting in the honeycomb
structure in Fig.~\ref{fig:Figure-ChargeAnaliz} can be taken as an indication
for the stability of 2D BN, calculation of phonon dispersion curves
through the diagonalization  of dynamical matrix provides a more
stringent test for stability. One of acoustical branches for
$\Gamma$ to $K$ curves taking negative value even at a small
region of BZ indicates the instability of the structure. There
have been a number of experimental \cite{Rokuta} and theoretical
studies of phonon spectrum of 2D \cite{Wirtz} and 3D honeycomb BN
\cite{Kern,Yu,Serrano,Solozhenkan, Miyamoto}. Here, the phonon
dispersion curves of h-BN, 2D BN and 1D BN chain and density of states together
with the infrared (IR) and Raman (R) active modes of 2D BN and h-BN at $\Gamma$-point 
have been calculated by using density
functional perturbation theory (DFPT) as implemented in PWSCF
software \cite{pwscf}. For the DFPT phonon calculation of bulk
h-BN, we used a four atom primitive cell, which yield 12 phonon
branches at the center of BZ in Fig. \ref{fig:Figure-Phonon} (a).
The symmetry point group is calculated as D$_{6h}$ (space group P6/mmm). The
irreducible representations at $\Gamma$ is 2 E$_{2g}$+ 2
B$_{2g}$+2 A$_{2u}$+2 E$_{1u}$. While the modes E$_{1u}$ and
E$_{2g}$ are doubly degenerate, B$_{2g}$ and A$_{2u}$ are non
degenerate. The modes E$_{1u}$ and  A$_{2u}$ are IR active, the
E$_{2g}$ is Raman active. B$_{2g}$ is an inactive mode. Our
results are in agreement with previously calculated and
experimental data, but differ slightly from those of Serrano et
al.\cite{Serrano}. While present GGA calculations predict
B$_{2g}$ mode as an inactive mode, LDA calculations by  Serrano et
al. found B$_{1g}$ as an inactive mode. Most of the phonon bands of
h-BN are degenerate. This indicates that the coupling between BN
layers in h-BN is weak. However, it is well known that the BN is
polar material with long range dipole-dipole interaction. This
gives rise to the splitting between longitudinal optical (LO) and
transverse optical (TO) mode  at $\Gamma$ point. The lowest
transverse acoustical mode has parabolic dispersion as \textbf{k}
$\rightarrow$ 0 owing to rapidly decaying interatomic forces for
transversal displacements \cite{decay}. Another feature is the
overlap of the lowest transversal optical mode with the acoustical
modes.

In Fig. \ref{fig:Figure-Phonon} (b) we show the phonon dispersion
curve of BN atomic chain. Two TA modes have low frequency and get
very small but negative values near the zone center. This
indicates structural instabilty as $\lambda \rightarrow \infty$. However, the
linear segments of BN atomic chain can be stable. Similar to h-BN,
the doubly degenerate TO branch overlaps with the LA branch.

For 2D BN honeycomb structure, the unit cell consists of two
atoms. Accordingly, there are three acoustical and three
optical branches in Fig. \ref{fig:Figure-Phonon} (c). The symmetry
point group is D$_{3h}$ (space group (P-62m)). Optical phonon
modes at the $\Gamma$-point is given by A$^{''}_2$+2 E$^{'}$. The
mode A$^{''}_2$ is IR active and the E$^{'}$ mode is both IR and
Raman active. The similarity between calculated phonon dispersion
curves of h-BN and 2D-BN is remarkable.

We also calculate the phonon dispersion curves of 2D BN honeycomb
structure by using PAW potentials \cite{paw} as implemented in
VASP \cite{vasp1} for further checking of the results of our
phonon calculation. Force constants are determined from the
$(8\times8\times1)$ supercells. The phonon modes were calculated
by using the direct method as implemented in the PHON \cite{alfe}
software. The calculated phonon frequencies are almost identical
with those calculated by DFPT method. In Fig.
\ref{fig:Figure-Phonon} (d), we present the phonon density of
states calculated for 2D BN honeycomb structure. Note that both calculations yield that TA (or ZA) mode displaying parabolic dispersion gets negative frequencies as $k \rightarrow 0$.
similar to BN atomic chains, this indicates structural instability as $\lambda \rightarrow \infty$. 
Accordingly, finite size of 2D BN flakes are expected to have stable structure.

\section{1D BN Nanoribbons}

Similar to graphene \cite{graphene-nanoribbons}, two unique
orientation in 2D BN yield nanoribbons with uniform edges: These
are armchair (A-BNNR) and zigzag (Z-BNNR) nanoribbons. The profile
of the atomic configuration at both edges of the nanoribbon
determines their electronic and magnetic properties. The
properties can be modified by the passivation of dangling bond of
edge atoms by hydrogen. Because of their interesting electronic and
spintronic properties, BN nanoribbons are attractive nanostructures
for various device applications. Electronic properties of BN
nanoribbons have been investigated in recent papers
\cite{Guo,louie-nanoletter,barone}. Present study is 
complementary to previous studies.

\subsection{Electronic structure}

\begin{figure}
\begin{center}
\includegraphics[width=7.5cm]{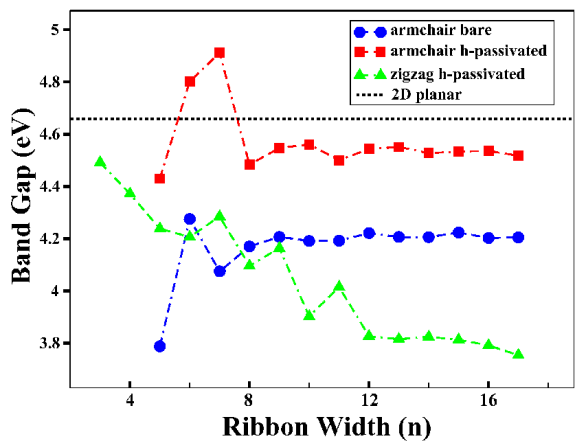}
\caption{(Color online). Energy band gap versus the width of the
nanoribbons given in terms of the number of B-N atom pairs in the
primitive unit cell, $n$. Bare armchair nanoribbons A-BNNR,
hydrogen passivated A-BNNR, and hydrogen passivated zigzag
nanoribbons Z-BNNR. Dotted line indicates the bulk band gap of
2D-BN. } \label{fig:Figure-BandGaps}
\end{center}
\end{figure}

Here we present the results of our study on the electronic and
magnetic properties of bare and hydrogen passivated A-BNNR and
Z-BNNRs. Bare and hydrogen passivation A-BNNR are wide band gap
semiconductors. Similarly, hydrogen passivated Z-BNNRs are also
semiconductor. The band gaps of these BN nanoribbons depend on the
width of the nanoribbons $w$ or the numbers of BN pairs, $n$ in
the primitive unit cell. The variation of the band gap E$_{G}$ as
a function of $n$ is given in Fig. \ref{fig:Figure-BandGaps}.
Normally, the properties of nanoribbons approaches to those of 2D
honeycomb structure as the width $n \rightarrow \infty$. However,
due to the localized edge states the band gap of Z-BNNR approaches
to a gap smaller than that of 2D BN honeycomb structure
\cite{louie-nanoletter}. For narrow ($n< 8$) bare and hydrogen
passivated A-BNNRs the band gaps vary with $n$, but they are
practically unaltered for $n>8$. For $n>8$ the band gap of bare
A-BNNR is 0.4 eV smaller than that of hydrogen passivated A-BNNR.
The band gap of hydrogen passivated Z-BNNR is 4.5 eV for $n=3$,
but decrease to 3.8 eV for $n=16$. However, its variation with $n$
is not monotonic for $5 < n < 13$, it rather display family
dependent oscillatory variation with  changes as large as 0.4 eV
between two consecutive values of $n$. On the other hand, bare Z-BNNRs are found to be
metallic.

\begin{figure}
\begin{center}
\includegraphics[width=7.5cm]{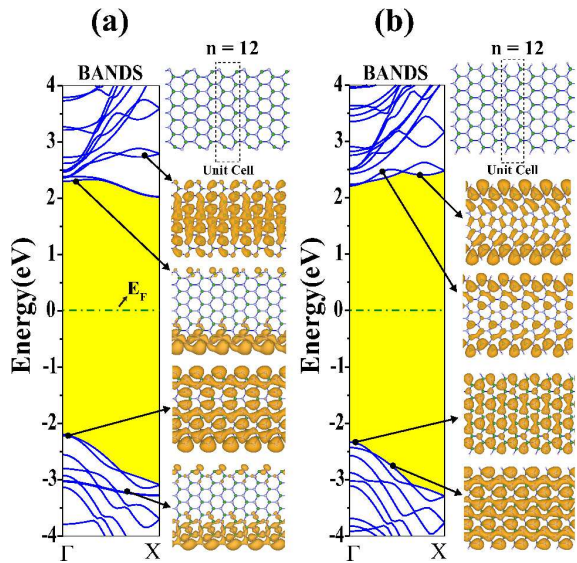}
\caption{(Color online) (a) Energy band structure of bare armchair
nanoribbon A-BNNR having $n=12$ B-N pairs in the primitive unit
cell. At the right hand side of bands, the schematic description
of atomic structure with primitive unit cell delineated by dotted
lines and isosurface charge distribution of specific states are shown. (b)
Same as (a) but the dangling bonds at both edges are passivated by
hydrogen atoms.} \label{fig:Figure-ArmChair}
\end{center}
\end{figure}

The atomic and electronic structure of bare and hydrogen
passivated A-BNNR are described in Fig.~\ref{fig:Figure-ArmChair}
for $n=12$. The atoms at the edges of the bare A-BNNR are
reconstructed; while one edge atom, B is lowering, adjacent edge
atom, N is raised. Two bands of edge states occur below the
conduction band edge. These bands are normally degenerate for
large $n$, but split around the center of BZ due to their coupling. The bands of edge states occur $\sim$-1 eV below the
top of the valance band edge. Normal states, on the other hand,
have charge distributed uniformly in the ribbon. Because of the
edge states the band gap is indirect and is $\sim$4.22 eV wide.
Upon passivation of the dangling bonds of B and N atoms situated
at the edges with hydrogen atoms, these edge state bands are
discarded from the band gap and reconstruction of edge atoms
disappear. At the end, the band gap of H-passivated A-BNNR becomes
direct and increases by $\sim$0.3 eV.

\begin{figure}
\begin{center}
\includegraphics[width=7.5cm]{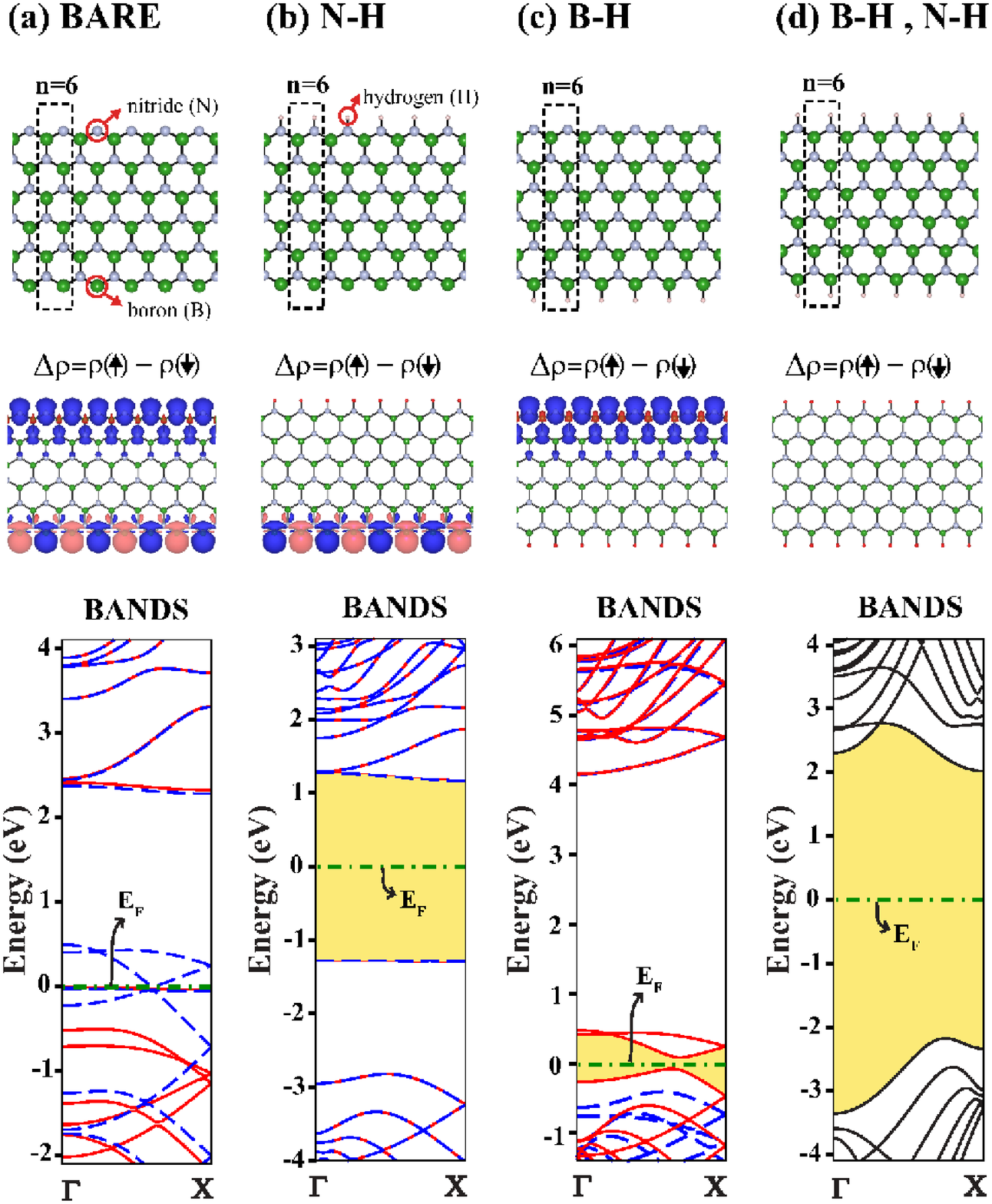}
\caption{(Color online) Top panels: Atomic structures of zigzag
nanoribbons (Z-BNNR). The primitive unit cell has $n=6$ B-N pairs
delineated by dotted lines. The unit cell is doubled due to
antiferromagnetic interaction between adjacent N atoms. Middle
panels: Isosurface plots of difference charge density between up
spin and down spin states, $\Delta \rho =
\rho(\uparrow)-\rho(\downarrow)$. Bottom panels: Energy band
structure with dotted (blue) and solid (red) lines showing spin up
and spin down states, respectively. (a) Bare Z-BNNR; (b) B-side
free, but N-side is passivated by hydrogen atoms; (c) N-side free,
but B-side is saturated by hydrogen atoms; (d) Both sides are
saturated by hydrogen atoms. The bands in (a), (b), and (c) are
calculated using double cell.} \label{fig:Figure-Zigzag}
\end{center}
\end{figure}

The electronic and magnetic states of Z-BNNR depend on whether
their edges are passivated with hydrogen atoms. While a bare
Z-BNNR is magnetic and metallic, it becomes nonmagnetic and a wide
band gap semiconductor upon the passivation of B and N atoms at
both edges. Moreover, its electronic and magnetic properties
depend on whether only B- or N-side is passivated with hydrogen
atoms. Accordingly, Z-BNNRs provide us for several alternatives
for different electronic and magnetic properties \cite{barone}.
However, different magnetic states corresponding to different edge
configuration, namely bare or hydrogen passivated, are very sensitive to the parameters of calculation.
In Fig.~\ref{fig:Figure-Zigzag} we present the calculated
electronic structures of a Z-BNNR with $n$=6 B-N pairs in a
primitive unit cell for four different cases. These are both side
free, only N-side is passivated with hydrogen, only B-side is
passivated with hydrogen and both edges are passivated with
hydrogen.

Bare Z-BNNR having both edges are free display different magnetic
states (magnetic order), which are close in energy. Moreover, the
ordering of these magnetic states with respect to their energy is
sensitive to the criterion of energy convergence.  To ensure the
antiferromagnetic (AFM) order at edges, we considered double cells. 
The possible magnetic states are spin-up,
spin-down for adjacent B atoms at one side and spin-up, spin-up
for the adjacent N atoms at the other side; namely
$\uparrow\downarrow$ / $\uparrow\uparrow$ spin configuration.
Other possible spin configurations are $\uparrow\uparrow$ / $\downarrow\downarrow$; $\uparrow\uparrow$ / $\uparrow\uparrow$
; $\uparrow\uparrow$ / $\uparrow\downarrow$; $\uparrow\downarrow$ / $\uparrow\uparrow$. We found that the spin configuration,
$\uparrow\downarrow$ / $\uparrow\uparrow$ for Z-BNNR having 12 B-N pairs in double unit cell corresponds to the
ground state. The other excited configurations, $\uparrow\uparrow$ / $\downarrow\downarrow$; $\uparrow\uparrow$ / $\uparrow\uparrow$
; $\uparrow\uparrow$ / $\uparrow\downarrow$; $\uparrow\downarrow$ / $\uparrow\downarrow$, have 6,7,35,131 meV  higher energies than ground state. The ordering of these configuration is slightly
different from that reported earlier \cite{barone}. Nevertheless,
the difference between the earlier and present ground state
energies are within the accuracy limits of DFT calculations. The
ground state spin configuration $\uparrow\downarrow$ /
$\uparrow\uparrow$ of the bare Z-BNNR having both edges free is
found to be ferrimagnetic metal with $\mu=1.77$ $\mu_{B}$ per
double cell. Whereas the excited
magnetic state with configuration $\uparrow\uparrow$ /
$\downarrow\downarrow$ is half metallic.

In Fig.~\ref{fig:Figure-Zigzag} (b), Z-BNNR with N-edge passivated
with hydrogen atoms is an AFM semiconductor. The AFM edge state is
localized at the B-side. When only the B-side is passivated with
hydrogen atoms the magnetic edge state is, this time, localized at
the N-side of the ribbon. As seen in Fig.~\ref{fig:Figure-Zigzag}
(c) the ground state of Z-BNNR is ferromagnetic with $\mu$ =2
$\mu_{B}$ per double cell. Our calculations
suggest that the nearest neighbor N-N interaction is
ferromagnetic, the B-B interaction is antiferromagnetic. Finally,
the Z-BNNR becomes non magnetic, when the atoms at both edges are
passivated with hydrogen atoms. Earlier, Hwan and
Louie \cite{louie-nanoletter} studied hydrogen passivated A-BNNRs
and Z-BNNRs with widths up to 10 nm. Our results for hydrogen
passivated nanoribbons are in good agreement with their results,
except that our results for zigzag ribbons obtained using GGA as
well as LDA exhibit family dependent oscillations for $5<n<13$.

\subsection{Elastic properties}

\begin{figure}
\begin{center}
\includegraphics[width=8.0cm]{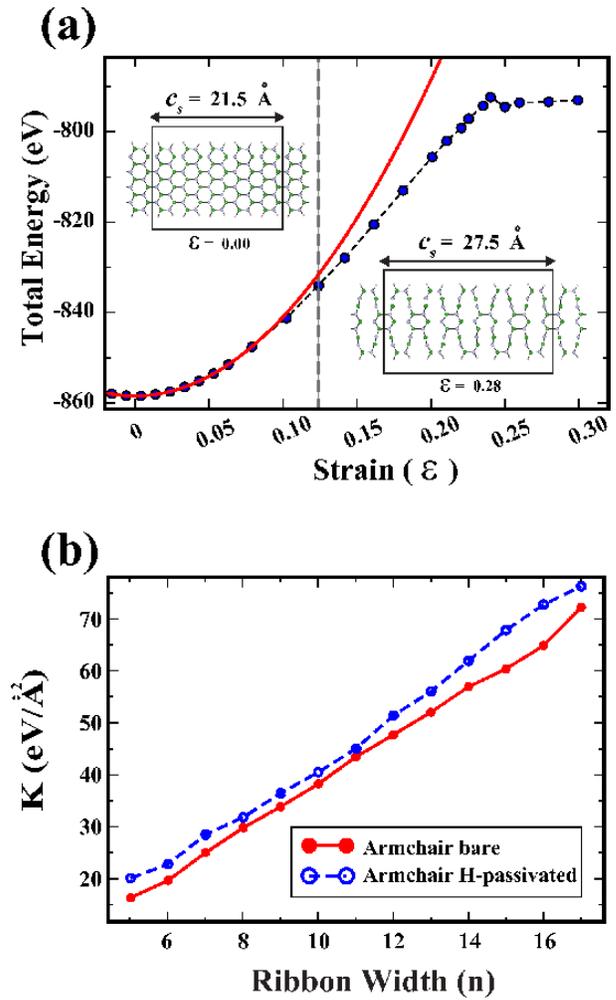}
\caption{(Color online) (a) Variation of total energy of hydrogen
saturated A-BNNR with strain , $\epsilon$ is shown by dashed curve
with large black dots indicating the calculated data points ($c_s = 5c$ and $n$ = 9). 
Harmonic, anharmonic and plastic regions
are distinguished. The harmonic part is fitted to a parabola
presented by red-solid curve. Atomic structure shown by filled,
empty and very small empty circles represent B, N, and H atoms.
Supercell comprising five primitive unit cells are shown in the
harmonic and plastic regions. (b) Variation of $\kappa =
\partial^2E_T/\partial c^2$ versus ribbon
width $n$ calculated for bare and hydrogen passivated A-BNNR.}
\label{fig:Figure-ArmChair-Kopma}
\end{center}
\end{figure}

The elastic properties of BNNRs are examined through the
variation of the total energy $E_{T}$ with respect to the applied
uniaxial strain $\epsilon = \Delta c/ c$, $c$ being the lattice
constant along the nanoribbon axis. Owing to ambiguities in
defining the cross section of the ribbon one cannot determine the
Young's modulus rigorously. Instead we calculate $\kappa =
\partial^2E_T/\partial c^2$ from the variation of $E_{T}$ to
specify the elastic properties of quasi 1D nanoribbons. In
Fig.~\ref{fig:Figure-ArmChair-Kopma} (a) we show the variation of
the total energy $E_{T}$ versus $\epsilon$. In order to lift the
constraints imposed by periodic boundary conditions, calculations
are performed for a supercell comprising five primitive unit cells
having lattice constant $c_s = 5c$. For $\epsilon < 0.10$,
the variation of $E_{T}(\epsilon)$ is parabolic, and hence
$\kappa$ is independent of $\epsilon$. For $\epsilon > 10$
$E_{T}(\epsilon)$ curve deviates from parabola and becomes
anharmonic. For higher values of strain in the plastic region, the
ribbon undergoes structural transformation. For example, such a
transformation occurred at $\epsilon=0.24$ with a sudden change in
$E_{T}(\epsilon)$ curve. The corresponding structure is
illustrated as inset. The lattice constant $c_s$ increased from
the initial value 21.5~\AA~to 27.4~\AA corresponding to
$\epsilon=0.27$.

In Fig.~\ref{fig:Figure-ArmChair-Kopma} (b) $\kappa$ versus the
width of the ribbon in terms of the number of B-N pair in the
primitive unit cell $n$ is plotted for bare and hydrogen
passivated A-BNNR. $\kappa (n)$ shows an approximately linear
variation indicating that the force constant is directly
proportional to the width of the ribbon. One also sees that the
strength of the ribbon increases upon passivation with hydrogen.

\begin{figure}
\begin{center}
\includegraphics[width=8.0cm]{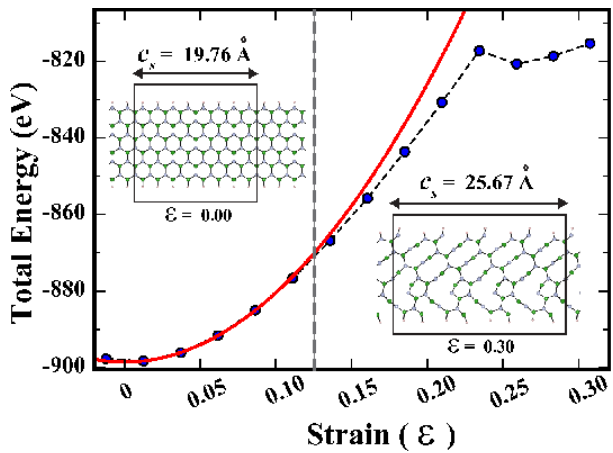}
\caption{(Color online) Variation of total energy of hydrogen
saturated Z-BNNR shown by dashed curve with large black dots
indicating the calculated data points. Harmonic, anharmonic and
plastic regions are distinguished. The harmonic part is fitted to
a parabola presented by a red solid curve. Atomic structure of the
ribbon in a supercell comprising eight unit cells ($c_s=8c$ and $n$ = 6) 
are shown before and after structural transformation as inset.}
\label{fig:Figure-Zigzag-Kopma}
\end{center}
\end{figure}

The behavior of bare and hydrogen passivated Z-BNNR under uniaxial
tensile stress is similar to that of A-BNNR. In
Fig.~\ref{fig:Figure-Zigzag-Kopma} three regions, namely elastic
harmonic, elastic-anharmonic and plastic regions are seen. The
sudden change in the $E_{T}(\epsilon)$ curve at $\epsilon \sim
0.23$ indicates a structural phase transformation, where the
lattice constant $c_s$ elongates from the initial $\epsilon = 0$
value of 19.8~\AA~ to 25.7~\AA~ corresponding to $\epsilon=0.3$.
The structure of hydrogen passivated Z-BNNR before and after the
structural transformation are shown as inset. Variation of
$\kappa$ versus the ribbon width $n$ is calculated for bare and
hydrogen passivated Z-BNNR show an overall linear behavior as
presented in Fig.~\ref{fig:Figure-ArmChair-Kopma} (b).

\subsection{Vacancy  and antisite defects}

It is known that the vacancy defect in 2D
graphene \cite{esquinazi,Iijima,yazyev,guinea,brey2} and graphene
nanoribbons \cite{brey1,delik} give rise to crucial changes in the
electronic and magnetic structure. According to Lieb's
theorem \cite{lieb}, the net magnetic moment per cell is determined
with the difference in the number of atoms belonging to different
sublattices, i.e. $\mu= (N_{B}-N_{N})\mu_{B}$. While DFT
calculations on vacancies in 2D graphene and armchair graphene
nanoribbons confirmed Lieb's theorem, results are diversified for
vacancies in zigzag graphene nanoribbons \cite{brey1,delik}.
Therefore, the effect of vacancy defects on the properties of
BNNRs is of interest.

Earlier activation energies and reaction paths for diffusion and
nucleation mono and divacancy in h-BN layers have been
investigated by using density functional tight-binding
method \cite{zobelli}. The formation energies were calculated to be
11.22 eV and 8.91 eV, for a B- and N-vacancy, respectively. The
possible magnetism induced by nonmagnetic impurities and vacancy
defects in a BN sheet have been investigated from the
first-principles. The magnetic moment associated by nonmagnetic
atoms substituting B or N has been calculated to be
$1\mu_{B}$ \cite{liu}. Based on first-principles calculations, the
magnetic moment of a N-vacancy in a 2D BN sheet has been predicted
to be 1 $\mu_{B}$. In the case of a B-vacancy, three neighboring N
atoms are displaced further apart from each other and the net
magnetic moment is predicted to be 3 $\mu_{B}$ \cite{si}. Another
calculation of defects in a BN monolayer found that three dangling
bonds associated with a B-vacancy lead to total spin S=3/2, i.e 3
$\mu_{B}$ \cite{azeveda}.

\begin{figure}
\begin{center}
\includegraphics[width=7.5cm]{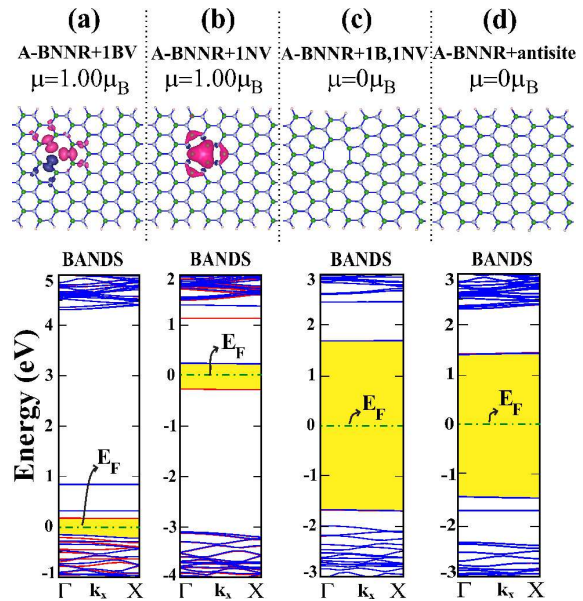}
\caption{(Color online) Relaxed atomic structures and
corresponding energy bands of hydrogen passivated A-BNNR with
$n=12$ having a point defect located periodically in every four
primitive cell. Blue filled, empty and small circles represent B,
N, and H atoms, respectively. Blue-dark and yellow-light
isosurface plots are for spin-up and spin-down states.  (a) Single
B-vacancy; (b) single N vacancy; (c) B-N divacancy; (d) Antisite
defect.} \label{fig:Figure-Vacancy-Armchair}
\end{center}
\end{figure}

The effects of vacancies of BN nanoribbons have not been treated
yet. Here we investigated the effect of B-, N-, B+N-divacancy and
B+N-anti site on the electronic and magnetic properties of A- and
Z-BNNR. Within periodic boundary conditions, a vacancy defect in
an A-BNNR of width $n=12$ is repeated in every 5th primitive unit
cell to yield minute defect-defect coupling. As shown in
Fig.~\ref{fig:Figure-Vacancy-Armchair} (a) A-BNNR with B-vacancy becomes
ferromagnetic with a net magnetic moment of $\mu=1$ $\mu_{B}$ per
unit cell. Similarly, a N vacancy gives rise to a net magnetic
moment of $\mu=1$ $\mu_{B}$ per unit cell. A-BNNR having either
periodic B+N-divacancy or anti site defect for every five unit
cell remain nonmagnetic. The calculated values of magnetic moments
are in compliance with Lieb's theorem. We found that the
structural relaxation is crucial to obtain correct values of
magnetic moments. In particular, initially we calculated $\mu=3$ $\mu_{B}$
for relaxed structure of the B-vacancy. However, the neighboring N atoms distorted slightly from their equilibrium, the structure is relaxed further and had lowered the total energy. As a result, the magnetic moment was calculated as $\mu=1$ $\mu_{B}$. The energy band structures in Fig.
\ref{fig:Figure-Vacancy-Armchair} (a)-(d) are calculated for
periodic vacancy defects repeating in every four primitive cell.
The Fermi levels are assigned according to the occupancy of
vacancy states. We note that the empty state associated with the
B-vacancy in Fig. \ref{fig:Figure-Vacancy-Armchair} is hole like.
The states associated with the N-vacancy occur near the edge of
the conduction band are donor like.
\begin{figure}
\begin{center}
\includegraphics[width=7.5cm]{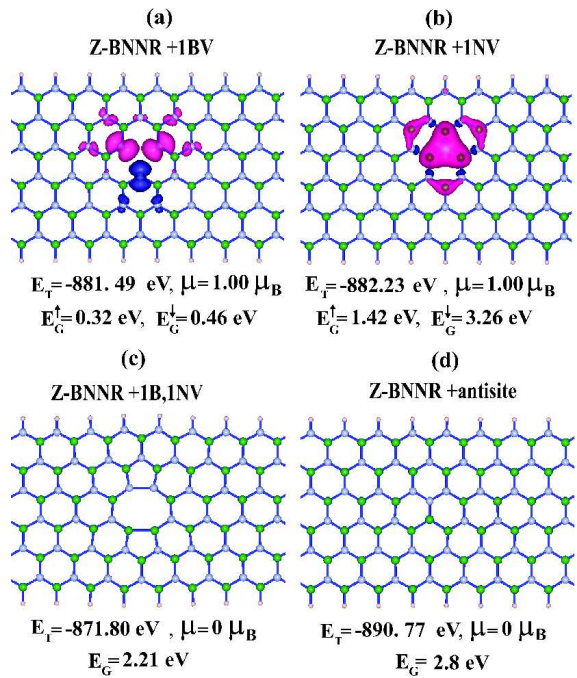}
\caption{(Color online) Relaxed atomic structures of hydrogen
passivated Z-BNNR with $n=6$ having a vacancy defect located
periodically in every eight primitive cell ($c_s \approx8c$ and
$n$ = 6). Filled, empty and small circles represent B, N, and H
atoms, respectively. Blue and pink isosurface plots are spin up
and spin down states, respectively. (a) Single B-vacancy; (b)
single N vacancy; (c) B+N divacancy; (d) anti site defect.}
\label{fig:Figure-Vacancy-Zigzag.eps}
\end{center}
\end{figure}

The situation with Z-BNNR is similar to that in A-BNNR discussed
above, since hydrogen passivated Z-BNNR is nonmagnetic as A-BNNR.
A periodic B- or N-vacancy repeated in every eight unit cell of
hydrogen saturated Z-BNNR with $n$=6 has a net magnetic moment of
$\mu=1$ $\mu_{B}$ per supercell. Whereas Z-BNNR passivated with
hydrogen atoms at both edges  and having either periodic
B+N-divacancy or anti site defect repeating in every eight unit
cell is nonmagnetic. The type of the periodic vacancy defect
modifies the band gap of Z-BNNR from 4.2 eV to 2.21 eV for
divacancy, but to 2.8 eV for anti site. The calculated magnetic
moment of hydrogen passivated Z-BNNR are in agreement with Lieb's theorem.  We note that in zigzag graphene nanoribbons magnetic edge states survive even after hydrogen passivation, and interact with the magnetic moments of vacancies \cite{delik}.
This interaction causes deviation from the prediction of Lieb's theorem.

\section{Discussion and Conclusions}
In various allotropic forms of BN the dimensionality play a
crucial role. For the sake of comparison, we present the
calculated values of BN for different allotropic forms of BN in
different dimensionalities. One sees that the B-N double bond of 1D
BN atomic chain is shortest and is 1.31~\AA. The s$p^2$ bond of
h-BN and 2D BN has intermediate value of 1.45~\AA. Therefore, h-BN
can considered to be quasi two dimensional. Three dimensional
wurtzite and zincblende BN crystal have s$p^3$-bonding with
$d=1.56$~\AA, which is largest among the allotropic forms studied
here. According to GGA results the cohesive energy of 2D BN is 3
meV larger than that of h-BN. This is due to fact that the GGA
calculation cannot account the van der Waals interaction between
atomic layers of h-BN. However, the calculations using LDA, where
the van der Waals interactions are better accounted, yield the
cohesive energy of h-BN is $\sim$57 meV larger than that of 2D BN
as one expects. The charge transfer $\Delta Q$ from B to N atom
increases with decreasing dimensionality. This due to fact that $d$
decreases with decreasing dimensionality. As for the coordination number 
increases with increasing dimension.

In 2D BN honeycomb structures and in its zigzag and armchair
nanoribbons, the B-N bond formed from the bonding s$p^{2}$ hybrid
orbitals from B and N atoms is essential. Owing to the transfer of
charge from B to N the B-N bond acquires an ionic character, which
underlies the semiconducting properties with wide band gap.

\begin{table}
\caption{Values of the bond length $d$  in \AA, cohesive energy
$E_C$ in eV per B-N pair, band gap E$_G$ in eV, charge transfer
from B to N ($\Delta Q$) in electrons and lattice constants (a,c) in \AA
calculated for various allotropic forms of BN in different
dimensionality.}\label{fig:table}
\begin{tabular}{|c|c|c|c|c|c|}
\hline
 & $d$ & $E_C$ & E$_G$ & $\Delta$ Q & Lattice \tabularnewline
\hline \hline 1D Chain & 1.307 & -16.04 & 3.99 & 0.511 & a=2.614
\tabularnewline \hline 2D BN & 1.452 & -17.65 & 4.64 & 0.429 &
a=2.511 \tabularnewline \hline h-BN & 1.450 & -17.65 & 4.47 &
0.416 & a=2.511 , c/a=2.66 \tabularnewline \hline Wurtzite & 1.561
& -17.45 & 5.726 & 0.342 & a=2.542 , c/a=1.63 \tabularnewline
\hline Zincblende & 1.568 & -17.49 & 4.50 & 0.334 & a=2.561
\tabularnewline \hline
\end{tabular}
\end{table}

Bare armchair nanoribbon of 2D BN is again a nonmagnetic wide band
gap semiconductor, the band gap of which is practically unaltered
with width $n>8$. Upon passivation with hydrogen band gap of the ribbon
increase by 0.3 eV. As for zigzag nanoribbons, they provide a
number of interesting properties. When its
both edges are bare, it is ferromagnetic metal. When its N-edge is
passivated with hydrogen, it becomes an antiferromagnetic
semiconductor. In the reverse case, namely when B-side is
passivated, it becomes a ferromagnetic semiconductor. When both
edges are passivated, it becomes a nonmagnetic, wide band gap
semiconductor. The band gap as well as the magnetic state of a
ribbon can be modified by periodic vacancy defects. Finally, BN
nanoribbons have been found to be strong, quasi one dimensional
and stable structures. They can sustain up to high strains, and
they stretch in the plastic region with  structural
transformations.

Briefly, the calculated electronic, magnetic and mechanical
properties of 2D BN honeycomb structure and its nanoribbons
present interesting but some differences from graphene. In this
respect BN honeycomb structure and its nanoribbons are
complimentary to graphene. The properties of 2D BN honeycomb
structure can be changed upon functionalization with foreign
atoms. Interesting quantum structures (such as single and series
quantum dots, resonant tunneling double barriers and multiple
quantum well structures) based on heterostructures and core shell
structures of lattice matched graphene and BN can be formed, since
the band gap of BN is much larger than that of graphene.

\begin{acknowledgments}
Part of the computational resources have been provided by UYBHM at
Istanbul Technical University through grant Grant No. 2-024-2007. E. Akt\"{u}rk gratefully acknowledgments the receipt of a BIDEB Postdoctoral Fellowship from TUBITAK.
\end{acknowledgments}


\begin{thebibliography}{99}

\bibitem{novo}
K. S. Novoselov,
A. K. Geim,
S. V. Morozov,
D. Jiang,
Y. Zhang,
S. V. Dubonos,
I. V. Grigorieva,
A. A. Firsov,
Science \textbf{306}, 666 (2004).


\bibitem{zhang}
Y. Zhang,
Y.-W. Tan,
H. L. Stormer,
P. Kim,
Nature \textbf{438}, 201 (2005).


\bibitem{berger}
C. Berger,
Z. Song,
X. Li,
X. Wu,
N. Brown,
C. Naud,
D. Mayou,
T. Li,
J. Hass,
A. N. Marchenkov,
E. H. Conrad,
P. N. First,
W. A. de Heer,
Science \textbf{312}, 1191 (2006).



\bibitem{graphene_applications1}
A. K. Geim, K. S. Novoselov, Nature Materials \textbf{6}, 183 (2007).

\bibitem{graphene_applications2}
Y.-W. Son, M.L. Cohen and S.G. Louie, Nature \textbf{444}, (2006).

\bibitem{graphene_applications3}
H. Sevin\c{c}li,
M. Topsakal,
E. Durgun,
S. Ciraci,
Phys. Rev. B \textbf{77}, 195434 (2008).

\bibitem{graphene_applications4}
M. Topsakal;
H. Sevin\c{c}li,
S. Ciraci,
Appl. Phys. Lett. \textbf{92}, 173118 (2008).

\bibitem{bn-synthesis}
K. S. Novoselov, D.  Jiang, F. Schedin, T. Booth , V. V. Khotkevich,
S. Morozov, A. K. Geim, Proc. Natl. Acad. Sci. U.S.A. \textbf{102}, 10451 (2005)

\bibitem{bn-nanosheets1}
D. Pacil\'{e}, J. C. Meyer, \c{C}. \"{O}. Girit, and A. Zettl, Appl. Phys. Lett. \textbf{92}, 133107 (2008).

\bibitem{bn-nanosheets2}
A. Nagashima, N. Tejima, Y. Gamou, T. Kawai, and
C. Oshima, Phys. Rev. Lett. \textbf{75}, 3918 (1995).

\bibitem{bn-nanocones}
L. Bourgeois, Y. Bando, W. Q. Han, and T. Sato, Phys. Rev. B
61, 7686 (2000)

\bibitem{bn-nanotubes}
 N. G. Chopra, R. J. Luyken, K. Cherrey, V. H. Crespi, M. L. Cohen,
S. G. Louie, A. Zettl, Science 1995, 269, 966.

\bibitem{bn-nanohorns}
C. Zhi, Y. Bando, C. Tang, and D. Golberg, Appl. Phys. Lett. 87, 063107 (2005).

\bibitem{bn-nanorods}
D. Golberg, A. Rode, Y. Bando, M. Mitome, E. Gamaly, and B.
Luther-Davies, Diamond Relat. Mater. 12, 1269 (2003).


\bibitem{bn-nanowires}
Y. J. Chen , H. Z. Zhang, Y. Chen, Nanotechnology \textbf{17}, 786 (2006).


\bibitem{Guo}
Z. Zhang and W. Guo, Phys. Rev. B \textbf{77}, 075403 (2008).

\bibitem{louie-nanoletter}
Cheol-Hwan Park and Steven G. Louie, Nano Lett. \textbf{8}, 2200 (2008).

\bibitem{barone}
V. Barone, J. E. Peralta, Nano Lett., \textbf{8}, 2210 (2008)

\bibitem{dai1}
X. Li,
L. Zhang,
S. Lee,
H. Dai,
Science \textbf{319}, 1229 (2008).

\bibitem{dai2}
X. Wang,
Y. Ouyang,
X. Li,
H. Wang,
J. Guo,
H. Dai,
Phys. Rev. Lett. \textbf{100}, 206803 (2008).


\bibitem{paw}
P. E. Bl\"ochl,
Phys. Rev. B \textbf{50}, 17953  (1994).


\bibitem{pw91} J. P. Perdew, J. A. Chevary, S. H. Vosko, K. A. Jackson, M. R. Pederson, D. J. Singh, C. Fiolhais, Phys.
Rev. B  \textbf{46}, 6671 (1992).

\bibitem{vasp1}
G. Kresse,
J. Hafner,
Phys. Rev. B \textbf{47}, 558 (1993).

\bibitem{vasp2}
G. Kresse,
J. Furthm\"{u}ller,
Phys. Rev. B \textbf{54}, 11169 (1996).

\bibitem{pwscf}
S. Baroni, A. Del Corso, S. Girancoli and P. Giannozzi,
http:/www.pwscf.org/

\bibitem{BNcrystal1}
A. Catellani, M. Posternak, A. Baldereschi, A. J. Freeman, Phys. Rev. B \textbf{36}, 6105 (1987).

\bibitem{BNcrystal2}
L. Liu, Y. P. Feng, and Z. X. Shen, Phys. Rev. B \textbf{68}, 104102 (2003).

\bibitem{BNcrystal3}
N. Ooi, V. Rajan, J. Gottlieb, Y. Catherine, and J. B. Adams, Modell. Simul. Mater. Sci. Eng. \textbf{14}, 515 (2006).

\bibitem{BNcrystal4}
K. Shimada, T. Sota, and K. Suzuki, J. Apl. Phys. \textbf{84}, 9 (1998).

\bibitem{BNcrystal5}
M. P. Surh, S. G. Louie, and M. L. Cohen, Phys. Rev. B \textbf{43}, 9126 (1991).

\bibitem{BNchain}
R.T. Senger, S. Tongay, E. Durgun and S. Ciraci, Phys. Rev. B
\textbf{72}, 075419 (2005).

\bibitem{bulk-references}
A.V. Kurdyumov, V.L. Solozhenko, and W.B. Zelyavski, J. Appl. Crystallogr. \textbf{28}, 540 (1995).

\bibitem{DFTgap}
Caution has to be taken in using the values of band gap calculated
with DFT as in the present study, since the value of band gap is
usually underestimated by DFT.

\bibitem{lowdin}
The calculations of $\Delta Q$ have been carried out by using PWSCF software \cite{pwscf}.

\bibitem{tongay}
S. Tongay, R.T. Senger, S. Dag and S. Ciraci, Phys. Rev. Lett.
\textbf{93}, 136404 (2004).

\bibitem{Rokuta}
E. Rokuta, Y. Hasegawa, K. Suzuki, Y. Gamou, C. Oshima and A. Nagashima, Phys. Rev. Lett. \textbf{79}, 4609 (1997).

\bibitem{Wirtz}
L. Wirtz, A.Rubio, R.A. delaConcha, A. Loiseau, Phys. Rev. B \textbf{68},045425 (2003).

\bibitem{Kern}
G. Kern, G. Kresse and J. Hafner, Phys. Rev. B \textbf{59}, 8551 (1999).

\bibitem{Yu}
W.J. Yu, W.M. Lau, S.P.Chan, Z.F. Liu, Q.Q. Zheng, Phys. Rev. B \textbf{67}, 014108 (2003).

\bibitem{Serrano}
J. Serrano, A. Bosak, R. Arenal, M. Krisch, K. Watanabe, T.
Taniguchi, H. Kanda, A. Rubio, and L. Wirtz,  Phys. Rev. Lett. \textbf{98},
095503 (2007).

\bibitem{Solozhenkan}
V. L. Solozhenkan, G. Will and F. Elf, Solid State Commun.
\textbf{96},1 (1995).

\bibitem{Miyamoto}
Y. Miyamoto, M. L. Cohen, and S. G. Louie,  Phys. Rev. B
\textbf{52}, 14971 (1995).

\bibitem{decay}
F. Liu, P. Ming and J. Li, Phys. Rev. \textbf{76}, 064120 (2007).

\bibitem{alfe}
D. Alf\`{e}, http://chianti.geol.ucl.ac.uk/~dario.

\bibitem{graphene-nanoribbons}
V. Barone, O. Hod and G.E. Scuseria, Nano Lett. \textbf{6}, 2748 (2006).

\bibitem{esquinazi}
P. Esquinazi , D. Spemann, R. H\"ohne, A. Setzer, K.-H. Han, T.
Butz, Phys. Rev. Lett. \textbf{91}, 227201 (2003).

\bibitem{Iijima}
A. Hashimoto,K. Suenaga, A. Gloter, K. Urita, S. Iijima, Nature
\textbf{430}, 870 (2004).

\bibitem{yazyev}
O. V. Yazyev and L. Helm, Phys. Rev. B \textbf{75}, 125408 (2007).

\bibitem{guinea}
M.A.H. Vozmediano, M.P. Lopez-Sancho, T. Stauber and F. Guinea,
Phys. Rev. B, \textbf{72}, 155121 (2005).

\bibitem{brey2}
L. Brey, H.A. Fertig, and S. Das Sarma, Phys. Rev. Lett.
\textbf{99}, 116802 (2007).

\bibitem{brey1}
J.J. Palacios, J. Fernandez-Rossier, and L. Brey, Phys. Rev. B.
\textbf{77}, 195428 (2008).

\bibitem{delik}
M. Topsakal, E. Akt\"{u}rk and S. Ciraci. Phys. Rev. B
\textbf{78},xxxxx (2008).

\bibitem{lieb}
E.H. Lieb, Phys. Rev. Lett. \textbf{62}, 1201 (1989).

\bibitem{zobelli}
A. Zobelli, C.P. Ewels, A.Gloter and G. Seifert, Phys. Rev. B
\textbf{75}, 094104 (2007).

\bibitem{liu}
R.F. Liu and C. Cheng, Phys. Rev. B \textbf{76}, 014405 (2007).

\bibitem{si}
M.S. Si and D.S. Xue, Phys. Rev. B \textbf{75}, 193409 (2007).

\bibitem{azeveda}
S. Azeveda, J.R. Kashny, C.M.C. de Castilhoand F. De Brito,
Nanotechnology \textbf{18}, 495707 (2007).


\end{thebibliography}
\end{document}